\pdfoutput=1
\documentclass[letterpaper,twoside,twocolumn,english,aps,prl]{revtex4}
\usepackage[T1]{fontenc}
\usepackage[latin9]{inputenc}
\usepackage{units}
\usepackage{amstext}
\usepackage{graphicx}
\usepackage{amssymb}
\usepackage{esint}

\makeatletter

\@ifundefined{textcolor}{}
{%
 \definecolor{BLACK}{gray}{0}
 \definecolor{WHITE}{gray}{1}
 \definecolor{RED}{rgb}{1,0,0}
 \definecolor{GREEN}{rgb}{0,1,0}
 \definecolor{BLUE}{rgb}{0,0,1}
 \definecolor{CYAN}{cmyk}{1,0,0,0}
 \definecolor{MAGENTA}{cmyk}{0,1,0,0}
 \definecolor{YELLOW}{cmyk}{0,0,1,0}
 }

\makeatother

\usepackage{babel}

\begin{document}

\preprint{This line only printed with preprint option}

\title{Observation of a cyclotron harmonic spike in microwave-induced resistances
in ultraclean GaAs/AlGaAs quantum wells }

\author{Yanhua Dai}

\affiliation{Department of Physics and Astronomy, Rice University, Houston, Texas
77251-1892, USA }

\author{R. R. Du}

\affiliation{Department of Physics and Astronomy, Rice University, Houston, Texas
77251-1892, USA }

\author{L. N. Pfeiffer}

\affiliation{Department of Electrical Engineering, Princeton University, Princeton,
New Jersey 08544, USA }

\author{K. W. West}

\affiliation{Department of Electrical Engineering, Princeton University, Princeton,
New Jersey 08544, USA }
\begin{abstract}
We report the observation of a colossal, narrow resistance peak that
arises in ultraclean (mobility $\sim3\times10^{7}cm^{2}/Vs$) GaAs/AlGaAs
quantum wells (QWs) under millimeterwave irradiation and a weak magnetic
field. Such a spike is superposed on the $2^{nd}$ harmonic microwave-induced
resistance oscillations (MIRO) but having an amplitude > 300\% of
the MIRO, and a typical FWHM $\sim$ 50 mK, comparable with the Landau
level width. Systematic studies show a correlation between the spike
and a pronounced negative magnetoresistance in these QWs, suggesting
a mechanism based on the interplay of strong scatterers and smooth
disorder. Alternatively, the spike may be interpreted as a manifestation
of quantum interference between the quadrupole resonance and the higher-order
cyclotron transition in well-separated Landau levels. 

PACS number(s): 73.43.Qt, 71.70.Di, 73.43.Jn
\end{abstract}
\maketitle
Magnetotransport in quantum Hall systems under an electromagnetic
wave has recently revealed unexpected new phenomena, including the
microwave-induced resistance oscillations (MIRO)\cite{1zudovprb2001,2Yeapl2001}
and zero-resistance states (ZRS) \cite{3maniprl2004,4Zudovprl2003}.
Such discoveries have stimulated much interest in the condensed matter
community \cite{5Dorozhkinjept2003,6willettprl2004,7Studenikinssc2004,8durstprl2003,9andreevprl2003,10dmitrievprb2005}.
Presently, the MIRO is being interpreted as resulting from either
a \textquotedblleft{}displacement\textquotedblright{} or a \textquotedblleft{}distribution\textquotedblright{}
mechanism \cite{8durstprl2003,9andreevprl2003,10dmitrievprb2005},
either of which could be responsible for a negative resistance under
proper conditions, leading to periodic oscillations in inverse magnetic
field, 1/B. In a very-high mobility two-dimensional electron system
(2DES) hosted in GaAs/AlGaAs heterostructures, the minima in MIRO
can reach ZRS. A microscopic mechanism for the formation of ZRS was
proposed in \cite{9andreevprl2003}, which invokes a spontaneous breaking
of translational symmetry and the formation of current or electrical
field domains. More recently, resistance oscillations and ZRS were
observed in a nondegenerate 2DES formed on the surface of helium \cite{11konstantinovprl2009}.
In this case transitions between Landau levels (LL) in different electrical
subbands are involved. Observations of ZRS in vastly different materials
systems underscore the fact that irradiated 2DES is a rich system
for studies of nonlinear transport where new phenomena continue to
emerge. 

In this Letter we report the observation of a colossal, narrow resistance
peak that arises in ultraclean (mobility $\sim3\ensuremath{\times}1\ensuremath{0^{7}}c\ensuremath{m^{2}}/Vs$)
GaAs/AlGaAs quantum wells (QWs) under millimeterwave (MW) irradiation
and a weak magnetic field. Such a spike is superposed on the $2^{nd}$
harmonic of the MIRO but having amplitudes > 300\% of the MIRO, and
a typical FWHM $\sim$ 50 mK. Such a photoconductivity (PC) peak does
not follow the \textquotedblleft{}phase shift\textquotedblright{}
pattern of MIRO \cite{1zudovprb2001,2Yeapl2001,3maniprl2004,4Zudovprl2003,5Dorozhkinjept2003,6willettprl2004,7Studenikinssc2004,8durstprl2003,9andreevprl2003,10dmitrievprb2005}
and represents a new effect in microwave-irradiated 2DES \cite{12yangprb2006}.
Further analysis of its frequency ($f_{MW}$)-dependence shows that
the spike occurs precisely at twice the cyclotron frequency, $2\pi f_{MW}/\mathrm{\omega_{C}=\omega/\omega_{C}=}2$,
where $\omega_{C}=eB/m^{*}$ and m{*} is the effective mass of electrons
in GaAs. Harmonics of the cyclotron resonance (CR) were previously
observed in far infrared (FIR) absorption experiments {[}13{]} and
theoretically interpreted in terms of the interplay of short range
scatterers and electron-electron interactions on low mobility Si MOFETs
{[}14{]}. In the high mobility GaAs/AlGaAs heterostructures, interaction
of collective excitations with CR harmonics has been reported {[}15{]}.
On the contrary, the PC, which is a dc response of the 2DES to electromagnetic
wave excitation, is generally known to occur as MIRO (not at the exact
CR harmonics). Systematic studies show a correlation between the spike
and a pronounced negative magnetoresistance (NMR), both observed in
our QWs, suggesting a mechanism based on the interplay of strong scatterers
and smooth disorder in very-high mobility, modulation-doped GaAs/AlGaAs
heterostructures {[}16{]}. 

\begin{figure}
\includegraphics[width=8.68cm]{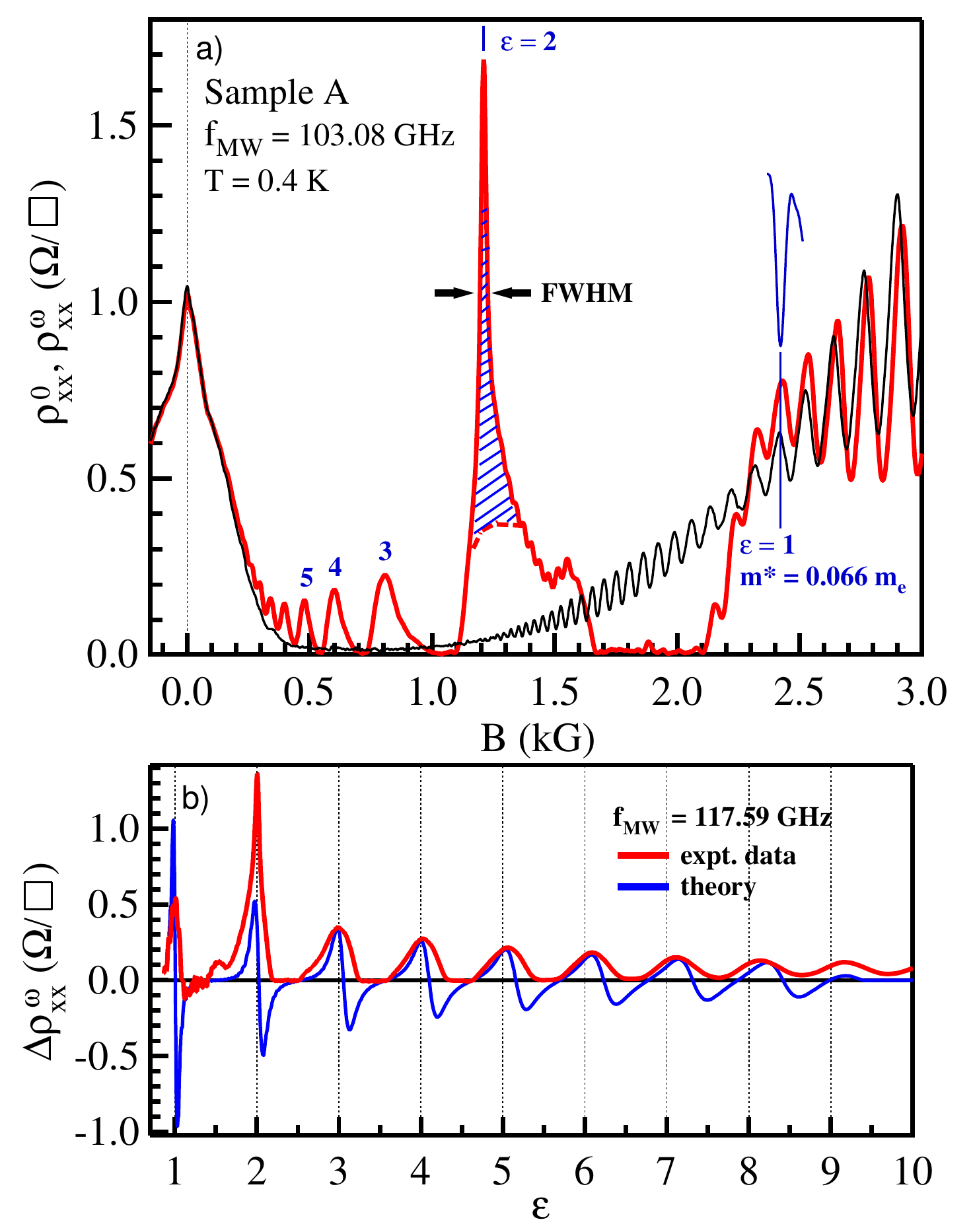}

\caption{\label{fig:overview1}(Color online) a) An example of the colossal
spike in magnetoresistivity $\rho_{xx}^{\omega}$ (hatched area) under
the millimeterwave irradiation is shown for sample A. The spike is
superposed on the MIRO. The trace $\rho_{xx}^{0}$ without irradiation
(black line) shows a strong NMR for magnetic field $B<1kG$. The inset
(blue line) is the second derivative of the bolometer signal, showing
a sharp minimum at cyclotron resonance. b) Photoresistivity $\Delta\rho_{xx}^{\omega}$
(red line) is plotted against the inverse magnetic field, 1/B, along
with the calculated MIRO trace (blue line).}

\end{figure}

Experimental results were obtained from 3 wafers of very-high mobility,
Si modulation-doped Al$_{x}$Ga$_{1-x}$As/GaAs /Al$_{x}$Ga$_{1-x}$As
QWs. All specimens were Hall bars (width between 90 and 180 $\mu m$)
defined by photolithography and wet etching; high quality electrical
contacts were made by Ge/Pd/Au alloy. Sample A, from which the main
data will be presented, has $x=0.24$, a well width $w=30nm$, and
is symmetrically doped with a spacer distance of $d=80nm$. After
a brief illumination from a red light-emitting diode the sample attained
an electron density of $n_{s}=2.9\times10^{11}/cm^{2}$ and a mobility
of $\mu\sim3\times10^{7}cm^{2}/Vs$ at $T=0.3K$. For comparison we
also took data from sample B, which has a very similar structure except
for that $w=25nm$. It has $n_{s}=4.6\times10^{11}/cm^{2}$ and $\mu\sim1.2\times10^{7}cm^{2}/Vs$.
Sample C, in which only the regular MIRO and ZRS, but not the spike,
were observed, has parameters $x=0.30$, $d=30nm$, $n_{s}=6\times10^{11}/cm^{2}$,
and $\mu\sim8.6\times10^{6}cm^{2}/Vs$. The experiments were performed
in a toploading $^{3}He$ refrigerator with a base temperature of
0.3K; experimental details can be found in {[}1, 4{]}. The magnetic
field was calibrated by a Gauss meter and all frequencies $f_{MW}<120\textrm{GHz}$
were calibrated by a frequency counter. An InSb bolometer was placed
directly behind a $3mm\times5mm$ piece of QW wafer (the same as for
sample A) for the CR experiments. 

An example of the PC is shown in FIG. 1a) with $f_{MW}=103\textrm{GHz}$
($\rho_{xx}^{\omega}$, red line); for comparison, the \textquotedblleft{}dark\textquotedblright{}
resistance $\rho_{xx}^{0}$ (black line) is also shown. The coolant
temperature for $\rho_{xx}^{0}$ was T = 0.32 K, whereas for $\rho_{xx}^{\omega}$
it rose slightly to 0.4K; the MW power incident on the sample is estimated
to be on the order of 100 $\mu W$, similar to the case of {[}4{]}.
We notice that the $\rho_{xx}^{0}$ exhibited a strong NMR with a
plateau minimum between $0.4kG<B<1kG$. The most prominent feature
in $\rho_{xx}^{\omega}$ is a spike (hatched area) at B$\sim$1.2
kG that has a magnitude as high as 300\% comparing to the MIRO on
the background. 

\begin{figure}
\includegraphics[width=8.68cm]{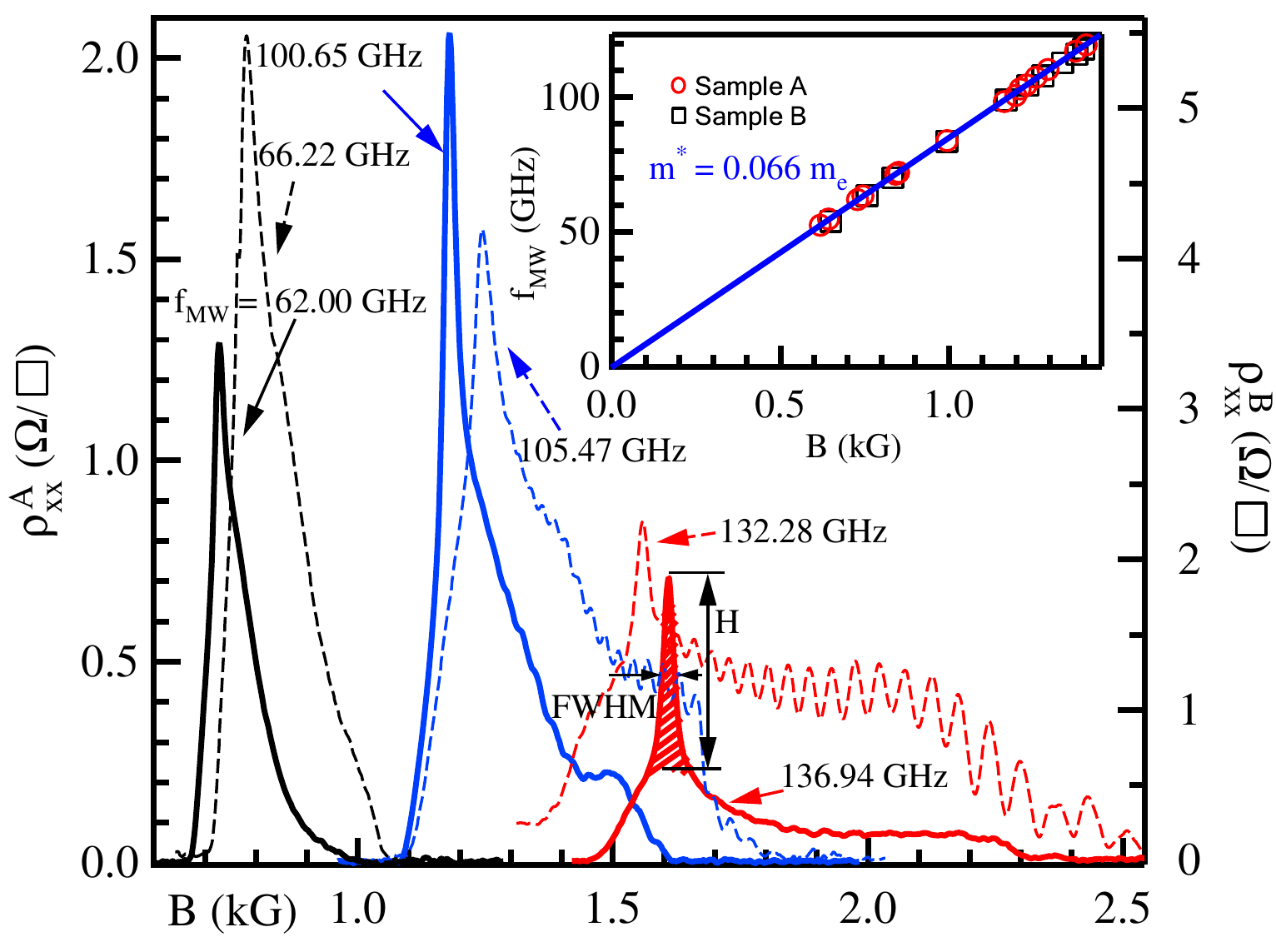}

\caption{\label{fig:frequency dependence2}(Color online) The frequency-dependence
of the spike is shown for respectively the sample A (solid lines)
and the sample B (dashed lines) in $f_{MW}$ between $\sim$60 and
$\sim$135 GHz. It is shown that the spike amplitude and width are
correlated with the sample mobility. The inset shows a linear relation
between the magnetic field position of the spike and the $f_{MW}$,
indicating that the spikes occur at $2\omega_{C}=2eB/m^{*}$, with
$m^{*}=0.066m_{e}$.}

\end{figure}

Cyclotron resonance was measured to determine the electron effective
mass; blue line in FIG.1a) shows the InSb bolometer signal $d^{2}R_{InSb}/dB^{2}$.
We observed a sharp minimum at $B=2.422kG$ corresponding to a CR
effective mass of $m^{*}=0.066m_{e}$, where $m_{e}$ is the free
electron mass. This value is within $\sim$1.5\% as compared to the
electron band mass in GaAs, $m_{b}=0.067m_{e}$. We then use this
value for calibration of $\varepsilon=\omega/\omega_{c}$. In particular,
we found that the spike position $B=1.210kG$, accurately yielding
$\omega=2\omega_{C}$ . For this reason, we refer to the spike described
here as the \textquotedblleft{}2$\omega_{C}$ peak\textquotedblright{}. 

In Fig. 1b) we plot $\Delta\rho_{xx}^{\omega}\equiv\rho_{xx}^{\omega}-\rho_{xx}^{0}$
vs. $1/B$ (red line) for $f_{MW}=117.59\textrm{GHz}$, where the
$\Delta\rho_{xx}^{\omega}$ is directly measured by a double-modulation
technique. $\Delta\rho_{xx}^{\omega}$ shows a series of ZRS up to
the $8^{th}$ order, attesting to the high quality of data. The $\Delta\rho_{xx}^{\omega}>0$
part, except for the $\varepsilon=2$ spike, is the well-known MIRO
showing a periodical pattern with a phase-shift $\delta\varphi$,
of which the value depends on the order of peak $j=\omega/\omega_{c}=1,2,3,\text{\ldots}$
\cite{17Zudov}. Specifically, $\delta\varphi$ tends to be close
to $\pi/4$ for higher orders but gradually diminishes towards the
major peaks $j=1,2$. The blue line is a fit to the MIRO \cite{7Studenikinssc2004}
by using \[
\textrm{\textrm{\ensuremath{\Delta\rho_{xx}}(B)=A\ensuremath{\int}d\ensuremath{\varepsilon}[\ensuremath{n_{F}}(\ensuremath{\varepsilon})-\ensuremath{n_{F}}(\ensuremath{\varepsilon}+\ensuremath{\hbar\omega})]\ensuremath{\nu}(\ensuremath{\varepsilon})\ensuremath{\partial_{\varepsilon}\nu}(\ensuremath{\varepsilon}+\ensuremath{\hbar\omega})}(Equ.1)}\]
Where A is a scaling factor for amplitude, $n_{F}(\varepsilon)=1/[1+\exp(\frac{\varepsilon-\varepsilon_{F}}{k_{B}T})]$
is the Fermi distribution function, and $\textrm{\ensuremath{\nu}(\ensuremath{\varepsilon})=\ensuremath{\sum\left(\frac{eB}{\pi^{2}\hbar\Gamma}\right)}/\ensuremath{\left\{  1+\left[\varepsilon-\left(i+\nicefrac{1}{2}\right)\hbar\omega_{c}\right]^{2}/\Gamma^{2}\right\} } }$is
the density of states with $\mathit{\Gamma}$ the LL broadening; $\varepsilon_{F}=15meV$
and $T=0.4K$ from the experiment. For $\varepsilon>2$ the calculated
magnetic field position and the amplitude of the MIRO fit the experimental
data quite well, yielding a LL line width $\Gamma\sim10\mu eV\sim120mK$.
Remarkably, a distinct 2$\omega_{C}$ peak is superposed on the MIRO
predicted by Equ. 1, indicating that it may be of a different origin.
As shown in FIG. 1 as well as in FIG. 2, the 2$\omega_{C}$ peak is
extremely narrow in width, and can be characterized by a large quality
factor $Q=B/\Delta B$ . For example, for $f=103\textrm{GHz}$ the
FWHM $\Delta B\sim0.026kG\sim50mK$ , leading to $Q\sim50$. A similar
procedure yielded $\mathit{\Gamma\sim200mK}$ and $\Delta B\sim100mK$
for sample B. Apparently, the $2\omega_{C}$ peak becomes prominent
for well-separated LLs, i.e., $\hbar\omega_{C}\gg\Gamma$, hence higher
$\mu$ or $f_{MW}$ favors the observation of this spike. 

\begin{figure}
\includegraphics[width=7.5cm]{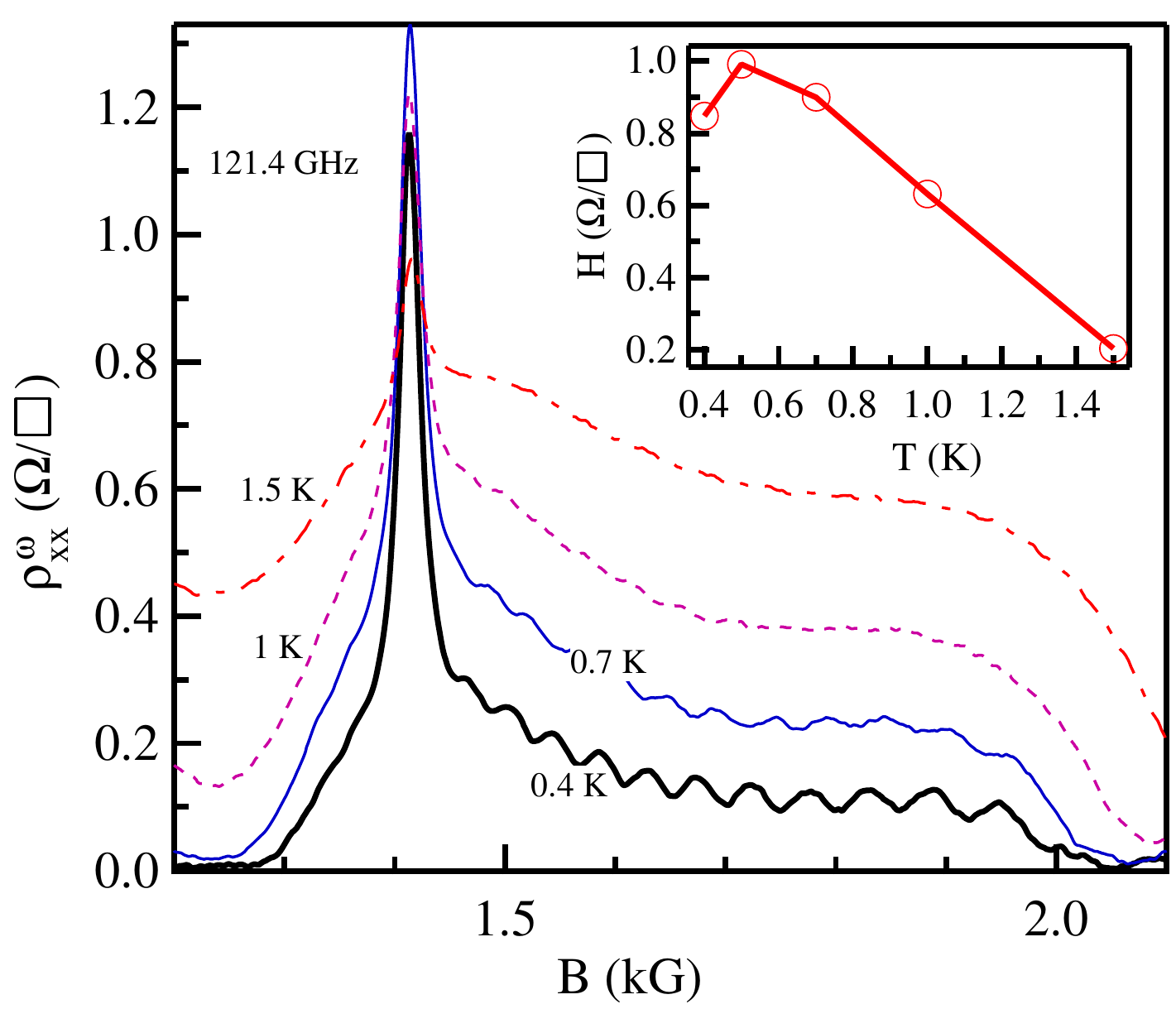}

\caption{\label{fig:TD3} (Color online) Spikes measured at different temperatures
are shown for the sample A. Inset: the spike amplitude shows an approximately
linear dependence with temperature. The line is a guide for the eye.}

\end{figure}

We show in FIG. 2 the 2$\omega_{C}$ peak position (in B) and its
width ($\Delta B$) in different MW frequencies, respectively observed
in the sample A and B. The observations can be summarized as follows:
1) The 2$\omega_{C}$ peak is a generic feature from a low frequency
$f_{MW}\sim60\textrm{GHz}$ to a high frequency $f_{MW}\sim135\textrm{GHz}$
in both samples; 2) The peak becomes more prominent as $\mathit{f}_{MW}$
increases; and, 3) The peak amplitude (as compared with the MIRO amplitude),
as well as its FWHM, is correlated with the sample mobility. In the
inset the peak position (in B) is plotted vs. $f_{MW}$, which shows,
again, that for both samples and in the whole MW range measured the
peak is associated with $\varepsilon=2$ with a fitted effective mass
$0.066m_{e}$. 

The amplitude of the 2$\omega_{C}$ peak shows a roughly linear dependence
on the coolant temperature. For example, in FIG. 3 inset we plot the
amplitude $H$ (defined as the total $\rho_{xx}^{\omega}$ subtracted
by MIRO) and found that the spike increased by a factor of 5 as T
decreased from 1.5 K to 0.5 K.

\begin{figure}
\includegraphics[width=8.68cm]{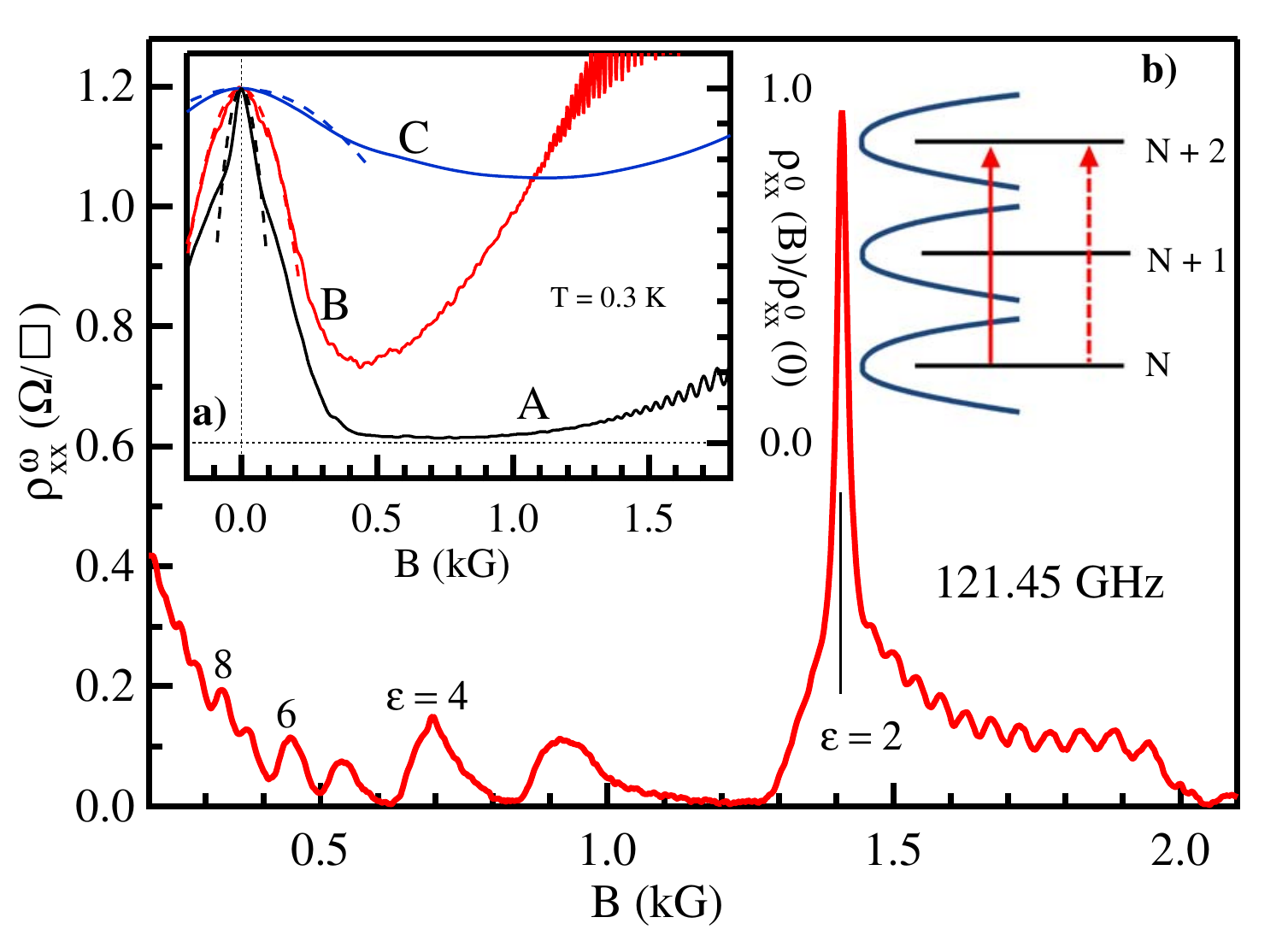}

\caption{\label{fig:illustration4} (Color online) Inset a): The 3 high-mobility,
modulation-doped GaAs/AlGaAs samples (A, B, C) show NMR, as explained
by a mechanism of interplay between strong scatterers and smooth disorder
\cite{16Mirlin}. The dashed lines are the asymptotic lines that yield
the characteristic frequency $\omega_{0}$ for the NMR (see the text).
In additions to the $\varepsilon=2$ spike, extra peaks can be observed
at the high-order even numbers of $\varepsilon$ ($\varepsilon=4,6,8$)
in the sample A. The inset b) is a schematic for cyclotron and quadrupole
transitions as discussed in the text. }

\end{figure}

In summary we have observed a new PC spike superposed on the regular
MIRO. Within the experimental accuracy, the spike is found to occur
precisely at $\varepsilon=2$. The finding is quite surprising, for
that its position, amplitude, as well as line-shape do not conform
to the descriptions of the presently accepted theoretical models,
neither the {}``displacement'' nor the {}``distribution'' mechanism.
Moreover, the spike can only be clearly observed under the conditions
of ultraclean GaAs/AlGaAs 2DES, more stringent than MIRO and ZRS.
In the following we discuss possible origins of the spike based on
its phenomenology.

In the Hall bar geometry pertaining to this experiment, magnetoplasma
(MP) with a wavevector $q=\pi/w$ can be excited by microwave, and
contribute to PC response in addition to MIRO {[}12{]}. In previous
work of FIR absorption in a grating-coupled 2DEG, Batke et al {[}15{]}
observed an interaction of the plasmon resonance with the second harmonic
of CR, and discussed the underlying mechanism in terms of nonlocality
of plasmon dispersion. While we cannot rule out the role of collective
plasma-like excitations (including the edge MP), we note that the
$2\omega_{C}$ peak can also be found in Corbino rings {[}4{]}. 

We focus now on the analysis of NMR, which are dominating features
in sample A and B (Fig. 4a)). Mirlin et al \cite{16Mirlin} considered
a two-component model of disorder in a very-high mobility, modulation-doped
GaAs/AlGaAs heterostructure containing: i) randomly distributed, dilute,
hard scatterers (termed {}``antidots'' here) with density $n_{S}$
and radius $a$ ($n_{S}^{-1/2}\gg a\gg k_{F}^{-1}$, where $k_{F}=\sqrt{2\pi n_{S}}$
is the Fermi vector), and ii) smooth random potential (correlation
radius $\sim d$, momentum relaxation rate $\tau_{L}^{-1}$, transport
mean free path $l_{L}=\nu_{F}\tau_{L}$, where $\nu_{F}=\hbar k_{F}/m^{*}$
is the Fermi velocity). The mean free path for the scattering on antidots
is $l_{S}=v_{F}$$\tau_{S}\approx1/2n_{S}a$. It is assumed that $\tau_{L}\gg\tau_{S}$,
so that the zero-B resistivity $\rho_{xx}^{0}(0)$ is determined by
antidots, $\tau^{-1}=\tau_{L}^{-1}+\tau_{S}^{-1}\approx\tau_{S}^{-1}$.
The combination of the two types of disorder induces a novel mechanism
leading to a strong NMR, followed by the saturation of $\rho_{xx}^{0}(B)$
at a value determined by the smooth disorder. 

As displayed in FIG. 4a), the sample A shows an unusually deep NMR
where the $\rho_{xx}^{0}(B)$ decreases by a factor of $\sim$50 at
B $\sim$ 0.5 kG and becomes a wide plateau. For sample B it shows
a steep valley at B $\sim$ 0.5 kG and then increases to a positive
magnetoresistance. Such behavior can be described consistently by
the above model. It is instructive here to estimate the relevant $n_{S}$
by a fit to the NMR. As shown by dashed lines (inset a) in FIG. 4),
the asymptotics \cite{16Mirlin} $\rho_{xx}(B)/\rho_{xx}(0)=1-(\omega_{C}/\omega_{0})^{2}$
describes reasonably well the onset of NMR $(\omega_{C}\ll\omega_{0})$,
where $\omega_{0}=(2\pi n_{S})^{1/2}\nu_{F}(2l_{S}/l_{L})^{1/4}$
is a characteristic frequency governed by the interplay of two scattering
components. Using the fitted values of $\omega_{0}$ and taking $l_{L}/l_{S}\sim50,10,5$
for A,B,C, we have determined the $n_{S}$ to be $(8\mu m)^{-2}$,
$(6\mu m)^{-2}$, $(2.6\mu m)^{-2}$, respectively. We conclude that
the 2D electrons in these samples experience scatterings by dilute
scatterers randomly distributed on a smooth background potential,
consistent with \cite{16Mirlin}. 

How the interplay of the two scattering components affects the photoconductivity
remains an interesting open question. Dimitriev et al \cite{18I. A. Dmitriev}
have studied theoretically this regime and predicted new features
in ac conductivity ($\Delta\sigma_{\omega}^{\left(C\right)}$) and
PC ($\sigma_{ph}^{(C)}$) beyond the standard MIRO. Briefly, the authors
address the non-Markovian corrections in the electron dynamics, which
were ignored in the Boltzmann treatment. They found an oscillatory
(in 1/B) correction $\Delta\sigma_{ph}^{\left(C\right)}\propto\Delta\sigma_{\omega}^{\left(C\right)}\propto-ReP\left(\omega\right)/n_{S}\tau_{S}$,
where the absorption $P\left(\omega\right)$ has a series of poles
at $\left(\omega-j\omega_{C}\right)/\Gamma$, j = 1, 2, 3, .... .
In principle such effect could be at the origin of the observed spike.
However, discrepancies exist, especially regarding the fact that we
have only seen a singular peak at $2\omega_{C}$ rather than oscillations. 

In addition, a mechanism based on quantum interference could play
an important role. Specifically, as depicted in the FIG. 4b), for
N to N+2 transitions there could exist two possible channels: i) due
to LL mixing the dipole transition between the N and N+2 levels (line
arrow), and ii) the quadrupole resonance (dashed arrow) in the presence
of a field gradient of millimeterwave. While interference between
the two channels was shown \cite{19A. P. Dmitriev} to generate photocurrent
at $2\omega_{C}$ in high B, its effect in very-high LLs has not been
addressed. Such interference effect, if confirmed by further experiments,
would be the evidence for {}``electromagnetically-induced transparency''
in the dc transport of an ac-driven 2DES as proposed in \cite{20Lee}. 
\begin{acknowledgments}
We thank Ivan Knez for the numerical simulation presented in FIG.
1b). We thank C. L. Yang for insightful discussion on the experiments
and especially on the fine points in data analysis. We acknowledge
I. A. Dmitriev and M. G. Vavilov for helpful communications, and thank
A. D. Mirlin, B. I. Shklovskii, S. A. Studenikin, and M. A. Zudov
for their interest and helpful comments. The work at Rice was supported
by NSF grant No. DMR-0706634. Use of Rice Shared Equipment Authority
for sample processing is acknowledged. \end{acknowledgments}

\end{document}